\documentclass[12pt]{article}
\usepackage{amssymb}
\usepackage{amsfonts}
\usepackage{latexsym}
\usepackage{amsmath}
\usepackage{amsthm}

\theoremstyle{plain}
\newtheorem{theorem}{\bf Theorem}[section]
\newtheorem{lemma}[theorem]{\bf Lemma}

\theoremstyle{remark}
\newtheorem{remark}[theorem]{\bf Remark}

\def\[{\begin{equation}}
\def\]{\end{equation}}

\newcommand{\x}[1]{x_{#1}}
\newcommand{\ve}{\varepsilon}
\renewcommand{\le}{\leqslant}
\renewcommand{\ge}{\geqslant}

\begin{document}

\title{Asymptotic Inverse Problem \\for Almost-Periodically Perturbed
\\ Quantum Harmonic Oscillator}
\author{
  Alexis Pokrovski
    \begin{footnote}
    {Laboratory of Quantum Networks, Institute for Physics,
    St-Petersburg State University, St.Petersburg~198504,
    Ulyanovskaya~1.
    E-mail: pokrovsk@AP15398.spb.edu}
    \end{footnote}
}
\date{}
\maketitle

\begin{abstract}
Let $\{\mu_n\}_{n=0}^\infty$ be the spectrum of
$-\frac{d^2}{dx^2}+x^2+q(x)$ in $L^2(\mathbb{R})$,
where $q$ is an even
almost-periodic complex-valued function
with bounded
primitive and derivative.
Suppose that the asymptotic approximation to the
 first asymptotic correction
$\Delta\mu_n=\mu_n-\mu_n^0+o(n^{-\frac{1}{4}})$,
$\mu_n^0=2n+1$ is known.
We prove the formula that gives the frequencies
and the Fourier coefficients of $q$
in terms of $\Delta\mu_n$.
\end{abstract}
81Q15 (Perturbation theories for operators
and differential equations).

\section{Introduction and main result}
Consider the operator describing perturbed quantum harmonic oscillator
\begin{equation}\label{MainOperator}
A=-\frac{d^2}{dx^2}+x^2+q(x) \qquad \textrm{in}\quad L^2(\mathbb{R})
\end{equation}
with the perturbation $q(x)$ from the class
$\mathcal{B} =\{q:\|q'\|_\infty+\|Q\|_\infty<\infty \}$,
 where $Q(x)=\int_0^xq\,dt$ and $\|\cdot \|_\infty$ denotes the norm in $L^\infty(\mathbb{R})$.
It was proved in \cite{ArXiv0} that the spectrum $\{\mu_n\}_{n=0}^\infty$ of $A$
has the asymptotics
$
\mu_n= \mu_n^0+\mu_n^1+O(n^{-\frac{1}{3}})$,
where
$\mu_n^0=2n+1$ and $\mu_n^1=O(n^{-\frac{1}{4}})$.

For the perturbations that are sum of almost-periodic and decaying terms
we study the problem of recovering of the almost-periodic
part from the first asymptotic correction $\mu_n^1$.
Specifically, we consider the perturbations
\begin{equation}\label{Decompq}
q=p+r\in \mathcal{B}, \quad   p\in B^1,p(-x)=p(x)\quad
\textrm{and}\quad
\|r\|_{B^1}\equiv
\lim_{T\to\infty}\frac{1}{2T}\int_{-T}^T |r(x)| dx=0,
\end{equation}
where $B^1$ is the Besikovitch space  of almost-periodic functions \cite{Besikovitch1954}
(closure of trigonometric polynomials
$\sum_{k=0}^N a_ke^{i\nu_kx}$, $\nu_k$ real,
 in the norm $\|\cdot\|_{B^1}$). It is sufficient to recover $p$
 in terms of its Fourier transform \cite{Besikovitch1954}.
Here is the main result.
\begin{theorem}\label{TheMainTheorem}
Let $\{\Delta\mu_n\}_{n=N}^\infty$
 approximates
the first asymptotic correction to the spectrum
of  the operator (\ref{MainOperator}),(\ref{Decompq}):

\begin{equation}\label{ApproximationToFirstCorrection}
\Delta\mu_n=\mu_n-\mu_n^0+o(n^{-\frac{1}{4}}).
\end{equation}

Then the spectrum and the Fourier coefficients of the
 almost-periodic part $p$  can be recovered from the relation
\begin{equation}\label{Spectrum-p}
\lim_{L\to\infty}
\frac{1}{\x{L}}\sum_{n=N}^{L-1}
\Delta\mu_n
G_\nu(\x{n},\x{L})
(\x{n+1}-\x{n})
=\lim_{T\to\infty}\frac{1}{T}\int_{0}^T p(t) \cos \nu t\, dt
,
\quad \nu\ge0,
\end{equation}
 where $x_n=\sqrt{\mu_n^0}=\sqrt{2n+1}$,
$
G_\nu(x,T)=-x\int_x^T
\frac{\varphi'_{\nu,T}(t) dt}{\sqrt{t^2-x^2}}$,
$
\varphi_{\nu,T}(t)=\eta(t-T)\cos\nu t
$
and   $\eta\in C^2(\mathbb{R})$ is a smoothed
step function such that $\eta(t)=1$ for $x\in(-\infty,-1]$,
$\eta(t)=0$ for $x\in[0,\infty)$ and $\eta'(0)=0$.
\end{theorem}

Asymptotic inverse spectral problem for quantum harmonic
oscillator with slowly decaying perturbation was considered by
Gurarie
 \cite{Gurarie}. He studied the operator (\ref{MainOperator})
 with real $q(x)\sim|x|^{-\alpha}{\sum a_m\cos\omega_m x}$
 for $|x|\to\infty$,
 where $\alpha>0$ and the sum in the numerator is finite.
The approach in \cite{Gurarie} is based on the spectral asymptotics
$$
\mu_n=\frac{\widetilde{q}(\sqrt{2n})}{n^{1/4+\alpha/2}}+
O(\frac{1}{\sqrt{n}}),
\qquad
\widetilde{q}(x)=\textrm{const}
\sum\frac{a_m}{\sqrt{\omega_m}}
\cos(\omega_m x-\pi/4)
$$
which exhibits linear relation between the leading asymptotic terms of
$q$ and $\mu_n$. However, the technique of \cite{Gurarie}
does not cover the case $\alpha=0$.

We consider just this case in a slightly more general setting
(almost-periodic  functions vs. finite trigonometric sums).
Our method also allows complex-valued $q$.
Technically, the result is based on the recent
proof \cite{ArXiv0} of the spectral asymptotics
\begin{equation}\label{FullAsymptotics}
 \mu_n=
\mu_n^0+
\frac{1}{2\pi}
\int_{-\pi}^\pi
q(\sqrt{\mu_n^0}\sin\vartheta)
d\vartheta
+
O(n^{-\frac{1}{3}}),
\qquad
\textrm{for} \quad
q\in \mathcal{B}.
\end{equation}
Thus the proof of Theorem~\ref{TheMainTheorem} follows from the
asymptotic behavior
 of the integral in
 (\ref{FullAsymptotics}), which is analyzed
  in Lemmas \ref{DirectSclomilch}
and  \ref{InverseSclomilch}.
\section{Properties of the Schl\"{o}milch integral}

The integral in the spectral asymptotics
 (\ref{FullAsymptotics}) is the
Schl\"{o}milch integral \cite{WhittakerWatson1}
\begin{equation}\label{DefSclomInt}
\qquad
g_q(x)=\frac{2}{\pi}
\int_{0}^{\pi/2}
q_+(x\sin\vartheta ) d\vartheta
=
\frac{2}{\pi}\int_0^x \frac{q_+(t) dt}{\sqrt{x^2-t^2}},
\qquad q_+(x)=(q(x)+q(-x))/2
\end{equation}
 evaluated at the points $x_n=\sqrt{2n+1}$.
In the next Lemma we estimate the integral and its
derivatives. Then in Lemma~\ref{InverseSclomilch}
we prove similar estimates for the inverse Schl\"{o}milch integral.
(We could not find in the literature the  results
of these Lemmas for the specific class $\mathcal{B}$.)
Using the two Lemmas,  we prove Theorem~\ref{TheMainTheorem}.

Everywhere below $C$ denotes an absolute constant.
\begin{lemma}\label{DirectSclomilch}
Let $f\in\mathcal{B}$ and $g(x)=\int_{0}^{\pi/2}
f(x\sin\vartheta ) d\vartheta$.
 Then
\begin{equation}\label{EstDirSchl}
|g(x)|\le C\frac{\|F\|_\infty+\|f\|_\infty}{\sqrt{1+x}},
\qquad
|g'(x)|\le C\frac{\|f\|_\infty+\|f'\|_\infty}{\sqrt{1+x}},
\quad
x>0,
\end{equation}
where $F(x)=\int_0^xf\,dt$.
\end{lemma}

\textbf{Proof.} For $x\le 1$ the result is evident, so we
consider only the case $x>1$.
 Using the change of variables
$t=\sin\vartheta$, we write
$g(x)=\int_0^1\frac{f(xt) dt}{\sqrt{1-t^2}}$ and
split it  as
\begin{equation}\label{SplitDirectSchl}
g=I_1+I_2, \qquad
I_1=\int_0^{1-\varepsilon} \frac{f(xt) dt}{\sqrt{1-t^2}},
\quad
I_2=\int_{1-\varepsilon}^1 \frac{f(xt) dt}{\sqrt{1-t^2}},
\end{equation}
where $\ve=1/x$.
Using the notation $(\ldots)_t'$ for
$\frac{\partial}{\partial t}(\ldots)$ and choosing the primitive
$\widetilde{F}$ of $f$,
satisfying $\widetilde{F}(x(1-\ve))=0$, we have\\
$
I_1=
\frac{1}{x}\int\limits_0^{1-\varepsilon}
\left(
\widetilde{F}(xt)
\right)_t'
\frac{dt}{\sqrt{1-t^2}}
=
\frac{1}{x}
\left(
\frac{\widetilde{F}(xt)}{\sqrt{1-t^2}}\Big|_{t=0}^{t=1-\ve}
+
\int\limits_0^{1-\varepsilon}
\frac{\widetilde{F}(xt) t\, dt}{(1-t^2)^{3/2}}
\right)
$. Therefore, 
\begin{equation}\label{EstI-1}
|I_1|\le
2\frac{\|F\|_\infty}{x}
\left(
1+
\int\limits_\ve^1\frac{dt}{t^{3/2}}
\right)
\le
C\frac{\|F\|_\infty}{x}
\left(
1+
\frac{1}{\sqrt{\ve}}
\right).
\end{equation}
Now we substitute the estimate
$|I_2|\le \|f\|_\infty\int_0^\ve\frac{dt}{\sqrt{t}}
=2\|f\|_\infty/\sqrt{x}$   and
(\ref{EstI-1}) into (\ref{SplitDirectSchl}).
This gives the
first inequality in (\ref{EstDirSchl}).
We prove the second one in a similar way,
writing
\begin{equation}\label{SplitDirectSchlDerivative}
g'=I_1'+I_2', \qquad
I_1'=\int_0^{1-\varepsilon} \frac{tf'(xt) dt}{\sqrt{1-t^2}},
\quad
I_2'=\int_{1-\varepsilon}^1 \frac{tf'(xt) dt}{\sqrt{1-t^2}},
\qquad\ve=1/x.
\end{equation}
We integrate by parts in $I_1'$, choosing the primitive
$\widetilde{f}(xt)=f(xt)-f(x(1-\ve))$. This gives
$I_1'=-\frac{1}{x}\int_0^{1-\varepsilon}
\frac{\widetilde{f}(xt)\,dt}{(1-t^2)^{3/2}}$, hence
\begin{equation}\label{EstI-1Prime}
|I_1'|\le
C\frac{\|f\|_\infty}{x}
\left(
1+
\frac{1}{\sqrt{\ve}}
\right).
\end{equation}
We substitute the estimate
$|I_2'|\le C\|f'\|_\infty/\sqrt{x}$  and
(\ref{EstI-1Prime}) in (\ref{SplitDirectSchlDerivative}).
 This gives the
second inequality in (\ref{EstDirSchl}). $\blacksquare$
\begin{remark}
The rate of decay $x^{-1/2}$ as $x\to\infty$ in
(\ref{EstDirSchl}) cannot be improved.
(The example $f(x)=\cos x$ gives the Bessel function $J_0$.)
\end{remark}

\begin{lemma}\label{InverseSclomilch}
Let $T>2$,
$\varphi,\varphi''\in L^\infty([0,T])$
and
$\varphi(T) =0$.
Then the equation
$\varphi(t)=\frac{2}{\pi}\int_t^T\frac{g(x) dx}{\sqrt{x^2-t^2}}$
has the unique solution
$g(x)=-x\int_x^T\frac{\varphi'(t) dt}{\sqrt{t^2-x^2}}$ for
$x\in[0,T]$, such that
\begin{equation}\label{EstInverseSchlomilch}
|g(x)|\le C(\|\varphi\|_\infty+\|\varphi'\|_\infty)
\sqrt{x},
\qquad \textrm{for} \quad x>1.
\end{equation}
If, in addition, $\varphi'(T)=0$, then
\begin{equation}\label{EstInverseSchlomilchPrime}
|g'(x)|\le C(\|\varphi'\|_\infty+\|\varphi''\|_\infty)
\sqrt{x},
\qquad \textrm{for} \quad
x>1.
\end{equation}
\end{lemma}
\textbf{Proof.}
In terms of
$\widetilde{g}(x)=\frac{g(\sqrt{x})}{2\sqrt{x}}$ and
$\widetilde{\varphi}(t)=\varphi(\sqrt{t})$
the equation on $g$ becomes  the Abel equation
$\widetilde{\varphi}(t)=
\frac{2}{\pi}
\int_t^{T^2}\frac{\widetilde{g}(s)\, ds}{\sqrt{s-t}} $.
Its solution for absolutely continuous
$\widetilde{\varphi}$  is
$
\widetilde{g}(s)=
\frac{\widetilde{\varphi}(T^2)}{\sqrt{T^2-s}}
-\int_s^{T^2}\frac{\widetilde{\varphi}'(u)du}{\sqrt{u-s}}
$
(see Ch.1, \S 2 of \cite{Samko}).
Using $\varphi(T)=0$, we obtain the required  formula for $g$.

Consider (\ref{EstInverseSchlomilch}).
For $ x\in [T-1,T]$ the inequality follows from the direct estimate
$|\frac{g(x)}{x}|\le
\frac{\|\varphi'\|_\infty}{\sqrt{2x}}
\int_{T-1}^T\frac{dt}{\sqrt{t-(T-1)}}$.
For $x\in [0, T-1]$ write
\begin{equation}\label{SplitInverseSchl}
 -\frac{g(x)}{x}=I_1+I_2,
 \qquad
 I_1=\int_x^{x+1}
 \frac{\varphi'(t) dt}{\sqrt{t^2-x^2}},
 \quad
I_2=\int_{x+1}^T
 \frac{\varphi'(t) dt}{\sqrt{t^2-x^2}}
\end{equation}
and integrate $I_2$ by parts. We have
$
I_2=
\frac{\varphi(t)}{\sqrt{t^2-x^2}}\Big|_{t=x+1}^{t=T}
-\int_{x+1}^T \varphi(t)
\tfrac{\partial}{\partial t}\frac{1}{\sqrt{t^2-x^2}}\, dt.
$
Therefore,
\begin{equation}\label{EstI2InvSchl}
|I_2|\le
\frac{\|\varphi\|_\infty}{\sqrt{1+2x}}
+\|\varphi\|_\infty
\frac{(-1)}{\sqrt{t^2-x^2}}\Big|_{t=x+1}^{t=\infty}
\le \frac{2\|\varphi\|_\infty}{\sqrt{1+2x}}.
\end{equation}
Now we substitute the estimate
$|I_1|\le\|\varphi'\|_\infty\int_x^{x+1}\frac{dt}{\sqrt{t^2-x^2}}
\le\|\varphi'\|_\infty\sqrt{2/x}$ and
(\ref{EstI2InvSchl}) into (\ref{SplitInverseSchl}). This gives
 (\ref{EstInverseSchlomilch}), as required.

Next consider (\ref{EstInverseSchlomilchPrime}).
By $g'(x)=x(g(x)/x)'+g(x)/x$ and
(\ref{EstInverseSchlomilch}), it is sufficient to estimate
 $x(g(x)/x)'$. Using $\varphi'(T)=0$, we obtain

\begin{equation}\label{CalcInvSchlomPrime}
x(g(x)/x)'=
-\frac{T\varphi'(T)}{\sqrt{T^2-x^2}}
+\int_1^{T/x}\frac{xs\varphi''(xs)\,ds}{\sqrt{s^2-1}}
=
\int_x^T\frac{t\varphi''(t)\, dt}{\sqrt{t^2-x^2}}.
\end{equation}
For $ x\in [T-1,T]$ we have
$|x(g(x)/x)'|\le
\frac{2x}{\sqrt{2x}}
\int_{T-1}^T\frac{|\varphi''(t)|\,dt}{\sqrt{t-(T-1)}}
\le 2\sqrt{2}\|\varphi''\|_\infty\sqrt{x}$.
For $x\in [0, T-1]$ write
\begin{equation}\label{SplitInverseSchlPrime}
 x(g(x)/x)'=I_1'+I_2',
 \qquad
 I_1'=\int_x^{x+1}
 \frac{t\varphi''(t) dt}{\sqrt{t^2-x^2}},
 \quad
I_2'=\int_{x+1}^T
 \frac{t\varphi''(t) dt}{\sqrt{t^2-x^2}}
\end{equation}
and take the integral for $I_2'$ by parts. We have
$
I_2'=
-\frac{(x+1)\varphi'(x+1)}{\sqrt{(x+1)^2-x^2}}
+
\int_{x+1}^T\varphi'(t)
\frac{\partial}{\partial t}\frac{t}{\sqrt{t^2-x^2}}\, dt.
$
Hence,  using
 $\frac{\partial}{\partial t}\frac{t}{\sqrt{t^2-x^2}}\le0$
 we obtain
 \begin{equation}\label{EstI2InvSchlPrime}
|I_2'|\le
\frac{(1+x)\|\varphi'\|_\infty}{\sqrt{1+2x}}+
\|\varphi'\|_\infty
\int_{x+1}^\infty\frac{\partial}{\partial t}\frac{(-t)}{\sqrt{t^2-x^2}}
\, dt
\le 2\|\varphi'\|_\infty\sqrt{1+x}.
\end{equation}
Now we substitute the estimate
$|I_1'|\le
\|\varphi''\|_\infty\int_x^{x+1}\frac{t dt}{\sqrt{t^2-x^2}}
\le
\|\varphi''\|_\infty\sqrt{1+2x}$ and
(\ref{EstI2InvSchlPrime}) into (\ref{SplitInverseSchlPrime}). This gives
 (\ref{EstInverseSchlomilchPrime}). $\blacksquare$

\begin{remark} Note that the condition
$\varphi(T)=0$ is necessary for (\ref{EstInverseSchlomilch}).
If $\varphi(T)\neq 0$, then $g(x)$ is unbounded due to
the non-integral term
in the inversion formula for the Abel equation.
The term is $O(\frac{\varphi(T)}{\sqrt{T-x}})$
for $x\uparrow T$. Similarly,
the condition $\varphi'(T)=0$ is necessary for
 (\ref{EstInverseSchlomilchPrime}).
If $\varphi'(T)\neq 0$, then $g'(x)$ is unbounded
as $x\uparrow T$ due to the non-integral term
$\frac{T\varphi'(T)}{\sqrt{T^2-x^2}}$
 in (\ref{CalcInvSchlomPrime}).
\end{remark}

\textbf{Proof of Theorem~\ref{TheMainTheorem}.}
Compare (\ref{ApproximationToFirstCorrection}) with
the asymptotis (\ref{FullAsymptotics}).
It is clear that
\begin{equation}\label{SchlomilchTransform}
\Delta\mu_n=g_q(\x{n})+o(n^{-\frac{1}{4}}),
\qquad
x_n=\sqrt{\mu_n^0}=\sqrt{2n+1},
\end{equation}
where the Schl\"{o}milch integral $g_q$ is given by
(\ref{DefSclomInt}).
The  proof is based on the
 fact that the set $\{\x{n}\}_{n=0}^\infty$
becomes arbitrarily dense as $n\to\infty$, so that
Riemann sums\\ $\frac{1}{T}\sum_{\x{n+1}\le T}
G_\nu(\x{n},T)g_q(\x{n})
(\x{n+1}-\x{n})$
approximate $\frac{1}{T}\int_0^{T}G_\nu(x,T)g_q(x) dx$,
provided the integ\-rand is smooth enough.
Using the inversion formula for
the Schl\"{o}milch integral,  we choose $G_\nu$ such that
 the last expression tends to
 $\frac{1}{T}\int_{0}^T q_+(t) \cos \nu t\, dt$ as
 $T\to\infty$.

By Lemma~\ref{InverseSclomilch},
we have
\begin{equation}\label{ApproxAverage}
\int_0^T
G_\nu(x,T)g_q(x)dx
=
\int_0^T q_+(t)
\left(
\frac{2}{\pi}\int_t^T\frac{G_\nu(x,T) dx}{\sqrt{x^2-t^2}}
\right)\,dt
=
\int_0^T
q_+(t)\varphi_{\nu,T}(t)
\, dt.
\end{equation}
Let us show that the integrand in the left-hand side
of (\ref{ApproxAverage}) is sufficiently smooth. By
Lemma~\ref{DirectSclomilch},
\begin{equation}\label{EstomatesOfg-q}
 |g_q(x)|\le
 C\frac{\|Q\|_\infty+\|q\|_\infty}{\sqrt{1+x}},
 \qquad
 |g_q'(x)|\le
 C\frac{\|q\|_\infty+\|q'\|_\infty}{\sqrt{1+x}},
 \qquad x\ge0,
\end{equation}
where $C$ is  an absolute constant.
 Similarly, since $\varphi_{\nu,T}$  satisfies the hypothesis of
  Lemma~\ref{InverseSclomilch}, uniformly in $T\ge x$
\begin{equation}\label{EstimatesOfG}
|G_\nu(x,T)|\le C(1+\nu)\sqrt{x},
\qquad
|\tfrac{\partial}{\partial x}G_\nu(x,T)|
\le
C(1+\nu)^2\sqrt{x},
\qquad x\ge1,
\end{equation}
where we used $\|\varphi_{\nu,T}\|_\infty\le C$,
$\|\varphi_{\nu,T}''\|_\infty\le C (1+\nu)^2$.
Therefore, for any fixed $\nu$ the function
$h_T(x)\overset{\text{\rm def}}{=}G_\nu(x,T)g_q(x)$
and its $x$-derivative
are uniformly  bounded for $x\ge1$ and $T\ge x$.
Hence, for $T\to\infty$ we have
\begin{equation}\label{ApproxIntG}
\frac{1}{T}\sum_{\x{n+1}\le T}h_T(x_n)(x_{n+1}-x_n)
-
\frac{1}{T}\int_0^{T}h_T(x)\,dx
=
\frac{1}{T}\int_0^{T}O(\frac{1}{x})\,dx
=
O(\frac{\ln T}{T})\to 0,
\end{equation}
where we used $\x{n+1}-\x{n}=O(\x{n}^{-1})$.
Now, by (\ref{ApproxIntG}), (\ref{SchlomilchTransform}) and the first estimate in (\ref{EstimatesOfG}),
\begin{equation}\label{ApproxDeltaMu}
\lim_{T\to\infty}
\frac{1}{T}\sum_{\x{n+1}\le T}
\Delta\mu_n
G_\nu(\x{n},T)
(\x{n+1}-\x{n})
=
\lim_{T\to\infty}
\frac{1}{T}\int_0^{T}G_\nu(x,T)g_q(x)\,dx.
\end{equation}
Next we divide (\ref{ApproxAverage})
  by $T$ and
 take the limit $T\to\infty$.
By (\ref{ApproxDeltaMu}) and
$
\lim\limits_{T\to\infty}
\frac{1}{T}\int_0^T
q_+ \varphi_{\nu,T}
\, dt
=
\lim\limits_{T\to\infty}\frac{1}{T}\int_{0}^T p(t) \cos \nu t\, dt
$,
this gives (\ref{Spectrum-p}).$\blacksquare$
\begin{remark}
We have
$(\nu x)^{-1}G_\nu(x,\infty)=
\frac{2}{\pi}
\int_x^\infty
\frac{\sin\nu t\, dt}{\sqrt{t^2-x^2}}=
J_0(\nu x)$, where $J_0$ is  the Bessel function
(see e.g. \cite{RyzhikGradstein}).
\end{remark}

\section{Acknowledgements}
The author is thankful to E.Korotyaev and S.Naboko
for fruitful discussions and valuable advice.

\end{document}